\def\cf{cooling flow}

\input aa.cmm
%\refereelayout
%
\MAINTITLE{Alfv\'en heating in optical filaments in cooling flows}
\AUTHOR={A.C.S Fria\c ca$^1$, D.R. Gon\c calves$^1$, 
L.C. Jafelice$^2$, V. Jatenco-Pereira$^1$, R. Opher$^1$}
\INSTITUTE={$^1$ Instituto Astron\^omico e Geof\'\i sico,
USP, Caixa Postal 9638, 01065-970, S\~ao Paulo -- SP, Brazil
\newline 
$^2$ Departamento de F\'\i sica Te\'orica e Experimental,
CCE-UFRN, Caixa Postal 1641, 59072-970, Natal -- RN, Brazil}
\ABSTRACT={
One of the major problems concerning cooling flows is the nature of
the mechanism powering the emission lines of the optical filaments
seen in the inner regions of cooling flows.
In this work we investigate the plausibility of Alfv\'en heating (AH) 
as a heating/excitation mechanism of optical filaments in cooling flows.
We use a time-dependent hydrodynamical code to follow 
the evolution of cooling condensations arising from the $10^7$ K cooling flow.
The filaments contain magnetic fields and 
AH is at work at several degrees of efficiency (including no AH at all).
We consider two damping mechanisms of Alfv\'en waves: nonlinear and turbulent.
We calculate the optical line emission associated
with the filaments and compare our results to the observations.
We find that AH can be an important ionizing and heating source
for class I filaments in the line-ratio scheme of Heckman et al.
In addition, AH can be an important contributor to [OI]$\lambda 6300$ emission
even for the more luminous class II systems.
}
\KEYWORDS={magnetic fields -- turbulence -- galaxies: clusters -- 
cooling flows -- intergalactic medium -- X-ray: galaxies}
\THESAURUS={02.13.1 -- 02.20.1 -- 11.13.1 -- 11.13.3 --  11.09.3 -- 13.25.2 }
\OFFPRINTS={A. C. S. Fria\c ca}
\DATE={(The date will be inserted later)}
\maketitle
\titlea{Introduction}
The short cooling times (less than a Hubble time) of the central
intracluster medium (ICM) of X-ray emitting clusters of galaxies points to
gas sinking towards the center of the cluster in a ``cooling flow"
(for reviews, see Fabian et al. 1984,1991 and Fabian 1994).
A sizeable proportion of the cooling flows also contains extended (from
$\la 1$ kpc to tens of kpc) optical emission line filaments around 
the central dominant galaxy of the cluster, which is always at the
center of the flow
(Heckman 1981; Cowie et al. 1983; Hu et al 1985; Johnstone et al. 1987;
Heckman et al. 1989; Baum 1992). 
These extended optical filaments systems are seen only in
cooling flows and are thought to be a phase of the evolution of thermal
instabilities arising in the cooling flow.
The line emission flux ratio
[NII]$\lambda$6583/H$\alpha$ allowed Heckman et al. (1989) to divide the
filaments into two distinct classes, class I with
an average $\langle$[NII]/H$\alpha \rangle = 2.0$, and class II with
$\langle$[NII]/H$\alpha \rangle = 0.9$.
Also, the class II filament systems tend to have
higher H$\alpha$ luminosity and to belong to
cooling flows with greater mass accretion rates.
More recently, however, the discovery of filaments in cooling flows
whose line ratios are intermediate between those typical
of class I and II systems (Crawford \& Fabian 1992;
Allen et al. 1992; Crawford et al. 1995)
called for a continuous distribution instead of a clear split into two classes.
In addition, there are some H$\alpha$-luminous
class I (i.e.  with high [NII]/H$\alpha$) systems
(e.g., A1068, A2146, RX J0439.0+0520) departing from the trend of class I
systems having low H$\alpha$ luminosities.
One of the major problems associated with the optical filaments is their
high luminosities (typically H$\alpha$ luminosities are $10^{41}$
to $10^{42}$ ergs s$^{-1}$).
Quiescently cooling gas that recombines only once
at the X-ray determined mass accretion rate would not be detectable as
emission line nebulae.
The fact that the central regions of cooling flows often do show
optical emission requires some source of energy to repeatedly reionize the gas.
Several models, considering different mechanisms for
ionization and heating of the gas,
-- shocks (David et al. 1988, David \& Bregman 1989),
thermal conduction (B\"ohringer \& Fabian 1989),
and photoionization by soft X-rays and EUV produced in the cooling gas
(Voit \& Donahue 1990; Donahue \& Voit 1991; Voit et al. 1994) --,
have met difficulties in explaining the observed line ratios and luminosities.
It is possible that some combination of photoionization and shock models
could explain the observations.
One example of these hybrid models involves the combination
of self-absorbed irradiating mixing layers of cold clouds embedded in the
hot cooling gas and emission from shocks generated in collisions between these
clouds (Crawford \& Fabian 1992).
Jafelice and Fria\c ca (1996) have investigated magnetic reconnection
as a heating mechanism and concluded 
that magnetic reconnection
cannot be invoked as the sole mechanism powering the optical
line emission in cooling flows. However, it could be an important ingredient
to explain the emission coming from low ionization lines. 
The models with high magnetic reconnection efficiencies 
exhibit a strong [OI]$\lambda$6300 emission,
and, therefore, magnetic reconnection could be an additional component 
together with other mechanisms producing most of the emission in H$\alpha$
and higher ionization lines (as [NII]$\lambda$6583).
The existence of magnetic fields in the ICM  has suggested magnetic reconnection
as a heating mechanism.
In this paper we investigate another heating mechanism
expected to be at work in the presence of magnetic fields:
Alfv\'en heating (AH). 
Here we explore the importance of AH not only as a dominant energy source
for optical
filaments in cooling flows but also as an ingredient contributing to 
the emission of some optical lines.
The Alfv\'en heating comes from the dissipation (damping) of Alfv\'en waves.  
These waves are easily generated in many cosmic plasmas, but they possess 
no linear damping mechanism since they are not compressive. Nonlinear 
damping of these waves occurs when one Alfv\'en wave decays into another 
plus a slow magnetosonic wave, or two Alfv\'en waves combine into  one fast 
magnetosonic wave; the resulting magnetosonic waves can then be 
dissipated. The dissipation of these waves may contribute to heat the solar 
corona, to influence the interplanetary wave spectrum, 
and to determine the wave spectrum available to scatter cosmic rays. 
As they are not compressive, the nonlinear 
decay occurs due to second order effects in the wave amplitude. Interacting 
Alfv\'en waves are compressive and thus generate a new compressive wave 
(see Chin and Wentzel 1972). The resulting heating --- at least in terms of 
nonlinear, resonance surface and turbulent damping mechanisms --- has been 
applied to study the thermal stability in the broad line region of quasars and 
in the basis of the wind in hot stars, in a successful way (Gon\c calves et 
al. 1993a, 1996a,b).   
Alfv\'en waves are created by perturbed magnetic fields.
In view of this, two conditions that are fulfilled by the core of a cooling flow
make it a favorable site for the generation of Alfv\'en waves:
it contains magnetic fields, and it is turbulent.
The lack of any strong hard X-ray emission due to inverse Compton
from the relativistic electrons in a few clusters 
(generally not cooling flows) that have halo radio sources
imposes a lower limit for the magnetic field of 0.1 $\mu$G
over scales of up to 500 kpc (Rephaeli \& Gruber 1988).
At the smaller ($\sim 100$ kpc) scales of cooling flows, magnetic fields
of typically $1-3$ $\mu$G are derived corresponding to 
a magnetic pressure of about 1\% of the thermal pressure.
At the center of a cooling flow still more intense magnetic fields 
must be present since the field is amplified by compression as the
gas flows inward (Soker \& Sarazin 1990).
As a matter of fact,
Faraday rotation and depolarization of the radio emission of extended 
radio sources (Ge \& Owen 1993; Taylor et al. 1994) residing in cooling flows
have revealed that the magnetic field increases inward and that it
can reach 20-100 $\mu$G depending on the degree of ordering of the field.
In this way, at the very center of the flow, the magnetic pressure can
supersede the thermal pressure.
One kinematic signature of the filament systems is the lack of
organized velocity patterns (e.g., rotation or shear or infall)
(Heckman et al. 1989; Baum 1992).
The filaments typically have very small rotational velocities and
are turbulently, and not rotationally, supported.
A small number of filament systems do show apparent rotation with rotational
velocities of $\sim 300$ km s$^{-1}$ within a few kpc. However, these same
systems show disordered velocity patterns at larger scales (Hu et al. 1985).
Lines are broad throughout the filamentary region, but line widths decrease
from 500-100 km s$^{-1}$ in or near the galactic nucleus to 100-300 km s$^{-1}$
at the largest radii (typically 5-15 kpc).
This suggests that the inner ICM is highly turbulent, with the amplitude
of random velocities increasing inward.
In fact, the core of a cluster is expected to be turbulent not only due to
the frequent crossing of galaxies through it but also due to relics
of merging of subclusters.
In addition, the absence of rotation in the center of cooling flows 
requires rotational breaking of the gas flowing inward,
and turbulent viscosity is the most likely transport process
of angular momentum in a cooling flow (Nulsen et al. 1984).
The possibility of using the turbulent energy to heat the plasma 
in the core of cooling flows was also considered by Loewenstein and Fabian 
(1990), in order to explain the kinematics of the cooling flow clouds.  
They assume a magnetic viscous heating process (` plasma slip') where 
the magnetic field that passes from the clouds to the hot gas can be forced 
to oscillate by the noise and so cause the ionized particles within 
the cloud to oscillate and collide with neutral particles. This process can 
efficiently transport Alfv\'en wave energy to length scales of $\sim 
10^{17}$ cm.
\titlea{Alfv\'en Heating in the Intracluster Medium}
The emission line filaments have line widths of several hundred to
$\sim$1000 km s$^{-1}$ (Heckman et al. 1989; Baum 1992; Crawford \& Fabian 1992;
Allen et al. 1992; Crawford et al. 1995). 
It is possible  that these velocities of the filaments
are turbulent and that the hot component of the ICM is itself turbulent.
Anyway, there is plenty of kinetic energy to be tapped via Alfv\'en waves
into an energy source for the ionization of the filaments.
Several damping mechanisms can be responsible for Alfv\'en heating.
In this work we consider the nonlinear damping 
and the turbulent damping for Alfv\'en waves. 
The damping mechanisms above have been investigated before in many 
astrophysical objects: late--type stars, protostellar and 
solar winds \ \ 
(Jatenco--Pereira \& Opher\- 1989a,b,c); galactic and 
extragalactic jets 
(Opher \& Pereira 1986; Gon\-\c cal\-ves et al. 1993b); 
early--type stars (dos Santos et al. 1993); 
broad line regions of quasars (Gon\c calves et al. 1993a, 1996a) 
and others.

\titleb{Nonlinear damping}
An Alfv\'en wave is likely to dissipate 
because of its nonlinear interaction with either the non--uniform 
ambient field or another Alfv\'en wave (Wentzel 1974). 
The nonlinear interaction of magnetohydrodynamic waves has been 
treated in detail by Kaburaki \& Uchida (1971), Chin \& Wentzel (1972), 
and  Uchida \& Kaburaki (1974). 
When the magnetic field is weak, i.e. when the Alfv\'en velocity $v_A$ 
is lower than the sound velocity $c_s$,
it is found that two Alfv\'en waves travelling in opposite directions 
along a magnetic field line can couple nonlinearly to give an 
acoustic wave, which in turn dissipates relatively quickly. 
In regions of strong ($v_A>c_s$) magnetic field, 
one Alfv\'en wave can decay into another 
Alfv\'en wave and a sound wave travelling in the opposite direction. 
The resulting Alfv\'en 
wave has a frequency smaller than the original one and it can, 
in turn, decay 
into another lower--frequency Alfv\'en wave plus an acoustic wave. 
The cascade continues until all the Alfvenic energy has been 
converted to acoustic waves that dissipate rapidly.
The latter possibility is considered in this paper.
In our models, the onset of nonlinear heating occurs for
$\beta = P_B/P_{gas} =(B^2/8\pi)/nk_B T  > \beta_{on} > 1$,
where $\beta_{on}$ is the initial value of $\beta$ when AH becomes important
($\beta>1$ corresponds to $v_A/c_s > \sqrt{6/5}$ for an ideal $\gamma=5/3$ gas).
Note that, following the convention widely used in cooling flow studies,
we define $\beta = P_B/P_{gas}$, whereas in plasma physics, 
$\beta_{pl}=P_{gas}/P_B$ denotes the ``beta parameter" of the plasma.
According to Lagage \& Cesarsky (1983), the nonlinear damping rate is 
$$
\Gamma_{NL} = {1 \over 4}\sqrt{\pi \over 2}\xi \bar {\omega}
\bigl({c_s \over v_A}\bigr) {\rho \langle \delta v^2 \rangle \over 
B^2/8\pi } \;\;, \eqno(1)
$$
\noindent where $\xi= 5 -10$, $c_s$ is the sound velocity, $v_A$  
the Alfv\'en velocity, $\bar {\omega}$  a characteristic Alfv\'en 
frequency 
and $\rho \langle \delta v^2 \rangle \big/(B^2/8\pi)$ 
the ratio of the energy density of Alfv\'en waves to that of the  
magnetic field.
As we are interested in AH coming from the nonlinear damping of 
Alfv\'en waves, we
heuristically derive the dependences of the heating rate  
$H_{NL}$ (erg cm$^{-3}$ s$^{-1}$) with 
density and temperature. 
Thus, the nonlinear Alfvenic heating is:
$$
H_{NL} = {\Phi_w \; \Gamma_{NL} \over 
v_A} \;\;, \eqno(2)
$$
\noindent where $\Phi_w$ is the wave flux,
$\Gamma_{NL}$ is given by (1), and $v_A= B /{\sqrt {4 \pi \rho}}$
is the Alfv\'en velocity. 
As our perturbation collapses, the component of the magnetic field
perpendicular to the direction of compression is amplified.
For our plane parallel geometry, $B \propto A^{-1} \propto \rho$,
where $A$ is the cross-sectional area perpendicular to the magnetic field.
As a consequence, $v_A \propto \rho^{1/2}$.
Let the dependence of $\Phi_w$  on $\rho$ be given as
$\Phi_w \propto \rho^c$.
Taking $\Phi_w = \rho\langle \delta v^2\rangle  v_A$, we have
$\rho\langle \delta v^2\rangle  \propto \rho^{c-1/2}$. 
Since $c_s \propto T^{1/2}$, for $\bar {w}$ independent of the density,
we obtain:
$$
H_{NL} \propto \rho^{2c - 7/2} \; T^{1/2} \;\;. \eqno(3)
$$
For a plane-parallel geometry, $\Phi_w \propto A^{-1} \propto \rho$, 
implying $c=1$, and equation (3) becomes:
$$
H_{NL} \propto \rho^{-3/2} \; T^{1/2} \;\;. \eqno(4)
$$
\titleb{Turbulent Damping}
\par\noindent 
Turbulence theory indicates that the spectrum of turbulent Alfv\'en 
waves has a power law dependence in phase space
$$ P(k) \propto k^{-\beta} \;\;\;, $$
\noindent where $k$ is the Alfv\'en wavenumber, $\beta$ is the spectral 
index, 
and 
$$ \int P(k)4\pi k^2 dk$$
is the energy density of turbulent Alfv\'en waves. The value of the spectral 
index is determined by the type of the turbulence (De Groot \& Katz 1973; 
Kainer et al. 1972; Eilek \& Henriksen 1984).
Tu et al. (1984) noted that the nonlinear interaction between outward 
and inward propagating waves results in an energy cascading process. 
A small part of the 
energy of lower frequency fluctuations cascades to higher frequency 
fluctuations until it reaches a frequency high enough for some dissipation 
process to occur, the most probable candidate being proton--cyclotron 
damping.
As noted by Hollweg (1987), the observed magnetic power
spectra, $P_{\rm B}$, in the solar wind exhibits a well-developed inertial
subrange resembling Kolmogorov turbulence, i.e. $P_{\rm B} \propto k^{-5/3}$,
over more than 3 orders of magnitude of $k$ (Matthaeus \& Goldstein 1982). 
This indicates a cascade to small scales.
The turbulent cascade transfers wave energy from large to short scales, 
and, in the end, the dissipation of energy through turbulence is governed by 
the small scale linear absorption mechanism.
Exploiting the similarity of
$P_{\rm B} \propto k^{-5/3}$ and Kolmogorov turbulence in ordinary fluids,
the volumetric heating rate associated with the cascade process
can be written as
$$
H_T=\rho{\langle \delta v^2 \rangle}^{3/2}/L_{\rm corr} \;\;\;, \eqno (5)
$$
\noindent
where $\rho$ is the mass density, $\langle \delta v^2 \rangle$ is the velocity 
variance associated with the field, and $L_{\rm corr}$
is a measure of the transverse correlation length (Holl\-weg 1986, 1987).
As in the previous subsection, we can derive the dependence
of the turbulent heating rate on $\rho$.
For a plane parallel geometry, 
$\langle \delta v^2 \rangle \propto \rho^{c-3/2}$
(assuming $\Phi_w \propto \rho^c$), 
and adopting $L_{\rm corr} \propto \rho^b$, we obtain:
$$
H_{T} \propto \rho^{(3/2)c -b - 5/4} \;\;. \eqno(6)
$$
For a plane parallel geometry, $c=1$ ($\Phi_w \propto \rho$),
and, if we assume that $L_{\rm corr} \propto B^{-1/2}$
(Holl\-weg 1986, 1987), then $b=-1/2$ (since $B \propto \rho$).
The resulting turbulent Alfv\'en wave heating is
$$
H_T \; \propto \; \rho^{3/4} \;\;\;. \eqno (7)
$$
\titlea{Time-dependent Models for Cooling Condensations}
We have investigated the evolution of optical filaments
since their formation out of the hot phase of the cooling flow
and calculated their optical signature
within the scenario outlined in Section 2.
The evolution of the cooling filaments is obtained by
solving the hydrodynamical equations of mass, momentum and
energy conservation (see Fria\c ca 1986,1993; Fria\c ca \& Terlevich 1994,1996).
The condensations are assumed to be self-gravitating.
Since there is no ionization equilibrium
for temperatures lower than $10^6$ K,
the ionization state of the gas at $T<10^6$ K is
obtained by solving the time-dependent ionization equations,
for all ionic species of H, He, C, N, O, Ne, Mg, Si, S, Ar and Fe.
We adopt an isochoric non-equilibrium cooling function for temperatures
lower than $10^6$ K, since the recombination time of important ions
is longer than the cooling time at these temperatures.
We have chosen an isochoric cooling function
because the evolution of the gas is nearly isochoric as soon as the gas
cools below $10^6$ K, first due to the rapid cooling for temperatures around
the peak ($T\sim 10^5$ K) of the cooling function, and then due
the magnetic support against further compression of the gas at lower
temperatures.
The cooling function and the coefficients of collisional ionization,
recombination and charge exchange of the ionization equations
are all calculated with the atomic database of the photoionization code
AANGABA (Gruenwald \& Viegas 1992).
The adopted abundances are half-solar,
as derived from X-ray spectroscopy of the ICM.
(Solar abundances are from Grevesse \& Anders 1989.)
The initial density perturbations
are characterized by an amplitude $A$ and a length scale $L$,
following $\delta\rho/\rho=A\,sin(x)/x$ (here $x = 2\pi\,r/L$).
We have also assumed a plane-parallel geometry and that the perturbations
are isobaric and nonlinear ($A$ = 1).
The geometry is justified by the fact that the line emission is filamentary,
which suggests that the perturbations are sheetlike rather than spherical.
We start to follow the evolution of the perturbations from the nonlinear
stage in view of the uncertainties about processes suppressing
the growth of thermal instabilities in cooling flows (see Section 5).
An unperturbed $n_H=0.1$ cm$^{-3}$ and $T=10^7$ K is assumed
for the filament, appropriate for the inner regions of
a cooling flow, where the filaments are more often seen.
$L$ was fixed at 1 kpc for all the models.
This length scale is suggested, for instance, by the spatial fluctuations in the
velocity of filaments resolved in nearby cooling flows (Heckman et al. 1989).
We fixed $\beta=P_B/P_{gas}=0.1$ for the unperturbed medium,
a representative value for the range of $\beta=0.01-1$, expected in the
central 10 kpc of cooling flows.  
\begtabfullwid
\tabcap{1}{Properties of models.}
\halign{
#\hfil&\quad\hfil#\hfil&\quad\hfil#\hfil&\quad\hfil#\hfil&\quad\hfil#\hfil
&\quad\hfil#\hfil&\quad\hfil#\hfil&\quad\hfil#\hfil&\quad\hfil#\hfil\cr
\noalign{\hrule\medskip}
\ \cr
Model & Heating & $\zeta$ & $\beta_{on}$
& $t_{col}$ & $t_{max}$ & $t_{1/2}$ 
& $\tilde L_{H\alpha,max}$ & $\tilde L_{H\alpha,ave}$ \cr
 & & & & (10$^6$ yr) & (10$^6$ yr) & (10$^6$ yr) 
& ($10^{40}$ erg s$^{-1}$) & ($10^{40}$ erg s$^{-1}$) \cr
\ \cr
A & none       & 0.  & --  & 6.00 & 0.345 & 4.51 & 2.65 & 1.08 \cr
B & turbulent  & 1.  & 1.  & 7.36 & 4.56  & 4.10 & 39.8 & 27.1 \cr
C & turbulent  & 0.3 & 1.  & 6.24 & 0.686 & 4.51 & 4.30 & 2.81 \cr
D & nonlinear & 0.3 & 3.  & 6.16 & 0.467 & 4.41 & 3.19 & 1.44 \cr
E & nonlinear & 1.  & 10. & 6.00 & 3.67  & 4.51 & 20.9 & 7.70 \cr
F & nonlinear & 0.3 & 10. & 6.00 & 0.447 & 4.51 & 3.20 & 1.40 \cr
\ \cr
\noalign{\hrule}}
\endtab
Here we investigate several representative models
with magnetic fields, one with no AH
and the others with AH for two damping mechanisms (nonlinear and turbulent)
and at different efficiencies (see Table 1).
Model A allows for the presence of a magnetic field only
through a magnetic pressure term ($P_B=B^2/8\pi$) in the equation of motion.
It gives the benchmark to evaluate the effects of AH.
In models B to F, the energy equation include AH as an additional
heating term, which is turned on only for $\beta > \beta_{on}$,
where $\beta_{on}$ is the value of $\beta$ when AH becomes important.
We considered nonlinear and turbulent Alfv\'en wave heating.
From equations (4) and (7), the heating term has the form
$$
H_{NL}=\zeta\,H_0 (n/n_0)^{-3/2}(T/T_0)^{1/2} \;\;, \eqno(8)
$$
\noindent
for nonlinear heating, and
$$
H_T=\zeta\,H_0 (n/n_0)^{3/4} \;\;, \eqno(9)
$$
\noindent
for turbulent heating,
where $\zeta$ is the efficiency of Alfv\'en heating.
The choice of the normalizations is the following:
1) $T_0=10^5$ K (the temperature at which optical emission begins to
be important);
2) $n_0=3.267\times 0.1$ cm$^{-3}$, where 3.267 is the compression factor
from the unperturbed $n_H=0.1$ cm$^{-3}$ and $T=10^7$ K initial state,
to $T=T_0$, under isobaric conditions;
and 3) $H_0=(3/2)\;n k_B T/t_{col}=2.55\times 10^{-24}$ erg s$^{-1}$ cm$^{-3}$,
where $t_{col}$ is the collapse time (defined below) for the filament
with no heating.
\titlea {Results}
\begfig 9cm \figure{1}
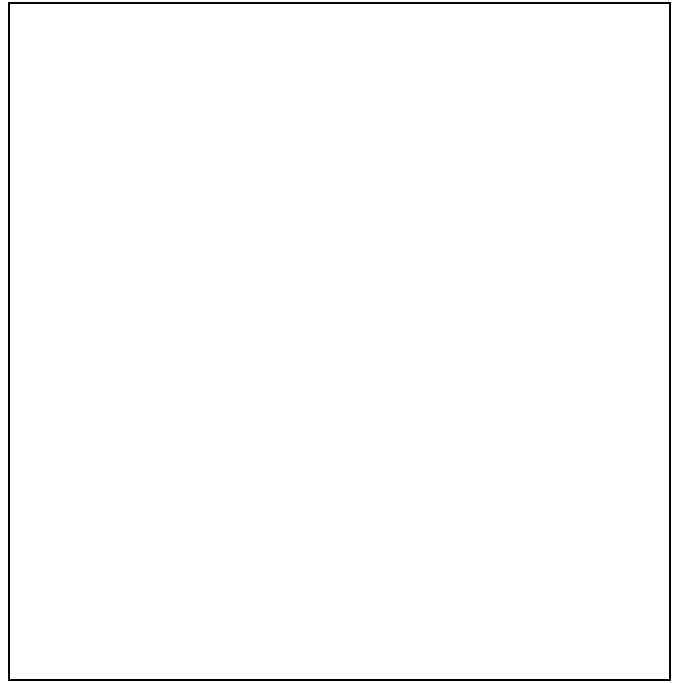
{Evolution of $\tilde L_{H\alpha}$, the normalized H$\alpha$ luminosity
(for its definition see the text),
for the models A-C (solid lines), D (dotted line) and E-F (dashed lines);
the curves are displaced upwards for increasing values of $\zeta$.}
\endfig
In order to assess the importance of the optical emission,
a number of quantities characterizing the optical phase of our models
are presented in Table~1:
the collapse time $t_{col}$ is the time elapsed since the
beginning of the calculations until the beginning of the optical phase;
$\tilde L_{H\alpha,max}$ is the maximum luminosity in H$\alpha$;
$t_{max}$ is the time when $\tilde L_{H\alpha,max}$ is attained;
$t_{1/2}$ is the time when half the mass of the perturbation has cooled
below $T=5\times 10^5$ K;
($t_{max}$ and $t_{1/2}$ are counted from the beginning of the optical phase);
and $\tilde L_{H\alpha,ave}$ is the average luminosity since
the beginning of the optical phase until $t_{1/2}$.
The luminosity $\tilde L_{H\alpha}$ is normalized to $\dot M$ = 100
M$_{\odot}$ yr$^{-1}$, with $\dot M$ taken as equal to the
ratio of the mass of the condensation to its collapse time.
In the beginning of the growth of the perturbation,
the magnetic pressure is unimportant in comparison to the thermal
pressure. However, as the gas within the perturbation cools,
under conditions of a frozen-in field, the component of the magnetic field
perpendicular to the direction of compression increases ($B \propto \rho$
in plane-parallel geometry), and eventually $\beta$ becomes $>1$.
At this stage, the magnetic pressure halts further compression, 
while the temperature keeps dropping (if no heating mechanism is
efficient enough to prevent the gas from cooling).
When the temperature has decreased enough, 
the gas begins to emit optical lines.
We consider the beginning of the phase of optical line emission
when the temperature of the innermost cell (i.e. the first cell
to cool) falls below $5\times 10^5$ K.
Before this temperature has been reached, the gas is in its X-ray phase,
during which most of the cooling occurs via EUV and X-ray lines.
The collapse time
is the smallest ($6\times 10^6$ yr) for models with no heating (model A)
or for those in which the heating is turned on later in the evolution
of the perturbation (models E and F).
As the heating becomes more important during the X-ray phase,
the collapse is delayed (i.e. $t_{col}$ is increased).
Model B (turbulent heating and $\zeta=1$) is the slowest to collapse.
A model with nonlinear heating, $\beta_{on}=3$, and $\zeta=1$, has been
run, but its mass that reaches $T< 5\times 10^5$ K is at most 35\% of
the total mass of the perturbation.
This result suggests that nonlinear Alfv\'en heating can constitute
a mechanism for suppressing the growth of thermal instabilities
in cooling flows.
The nonlinear and the turbulent AH's represent extreme opposite 
dependencies of AH on density.
Turbulent AH ($H_T \propto n^{3/4}$) deposits more energy 
at the center of the filament, where the density is higher.
On the other hand, nonlinear AH ($H_{NL} \propto n^{-3/2} T^{1/2}$)
deposits more energy per unit volume in the outer regions of the
filaments, where the density is lower.
However, due to magnetic support, the central density of the cooled
filament ($T\approx 10^4$ K) is only $\approx 3.3$ times the
initial unperturbed density,
and the central decrease of the AH rate due to density variation amounts
to only a factor 6.
In the end, the fact that the energy is deposited in a much cooler gas 
at the center
than on the outside of the filament ($T\approx 10^7$ K)
makes the AH somewhat more important in the colder, central gas
than in the outer skin of the filament,
as we can see from the timescale for heating
$t_H=(3/2)nk_BT/H_{NL} \propto n^{5/2}T^{1/2}$,
which is 1.6 times smaller at the center than on the outside of the filament.
The inverse dependency of $H_{NL}$ on the density explains the
suppression of the growth of the perturbation in the model 
with non-linear heating, $\zeta=1$ and $\beta_{on}=3$.
It should be noted that in our calculations, we assume that the filaments
are optically thin to Alfv\'en waves, so that energy is deposited by
AH throughout the filament.
The evolution of $\tilde L_{H\alpha}$ is shown in Figure~1.
and the evolution of the line ratios can be followed in the diagram
[NII]$\lambda$6583/H$\alpha$ versus [OI]$\lambda$6300/H$\alpha$ (Fig.~2).
For a model to be successful, it should
not only reproduce the position of the filaments in the line ratio diagrams
but also account for their luminosity.
The models B (turbulent heating) and E (nonlinear heating) are the more
energetically promising.
The energy budget requirements of class I filaments
are satisfied (see Table~1 and Fig.~1)
since the class I objects are the less luminous systems 
(an average $\langle \tilde L_{H\alpha}\rangle =4\times10^{40}$ erg s$^{-1}$).
Our models even fulfill the luminosity requirements 
for some of the less luminous class II filaments, with
an average $\langle \tilde L_{H\alpha}\rangle =3\times10^{41}$ erg s$^{-1}$
(Heckman et al. 1989).
The luminosity of our models, however, fall short of the most luminous class II
systems (e.g. Perseus), so that we propose the AH as the mechanism
powering primarily class I filaments.
From the 
[NII]/H$\alpha$ versus [OI]/H$\alpha$ diagram, the
model B (with turbulent heating) reproduces the values of [OI]/H$\alpha$
of the filaments, but somewhat overproduces [NII]$\lambda$6584 emission
of class I filaments.
On the other hand, model E (with nonlinear heating) reproduces the 
values of [NII]/H$\alpha$, but overproduces [OI]$\lambda$6300 emission.
It seems that if some combination of these mechanisms is at work,
a range of heating efficiencies and chemical abundances could
account for the loci of class I systems in the
[NII]/H$\alpha$ versus [OI]/H$\alpha$ diagram.
\begfig 9cm \figure{2}
{Evolution of [NII]/H$\alpha$ versus [OI]/H$\alpha$
for the models A-C (solid lines), D (dotted line) and E-F (dashed lines);
the curves are displaced upwards for increasing values of $\zeta$.
The plus signs on the curves indicate the elapsed times
(starting on the left) since the
beginning of the optical emission phase:
$2\times 10^{5}$, $2.5\times 10^{5}$, $3\times 10^{5}$ yr,
and from then on in intervals of $10^{5}$ yr until $4\times 10^{6}$ yr.
The contours enclose the data points of Heckman et al. (1989).
Note the clear separation between filaments of class I and II.}
\endfig
\titlea {Conclusions}
From the energetic viewpoint, our models could account for typical
filament systems with $H_{rec} \approx 10-100$ but not for the extreme ones
(e.g., Perseus) with $H_{rec} \approx 1000$ 
($H_{rec}$ is the number of recombinations per cooling proton required to
yield the observed H$\alpha$ luminosity).
Our models are better at reproducing the luminosities
of the weaker class I filaments than the more luminous class II systems.
AH could be a dominant heating mechanism for the class I filaments, and
even for the less luminous class II filaments.
It should be noted that
there are evidences that matter is dropping out of the cooling flows
at radii larger than $r\sim 10$ kpc, so that only a fraction of the 
overall \cf\ rate (over typical radii of $\sim 100$ kpc) 
reaches the region where the filaments are seen.
A fraction $\phi \approx 0.1$ reaching $r\sim 10$ kpc 
is suggested from a $\dot M \propto r$ variation
(Thomas, Fabian \& Nulsen 1987).
This value is derived from a simple model in which the flow is stationary
and the X-ray emission comes from a homogeneous gas.
Taking into account nonsteady flow 
(the flow time from 100 to 10 kpc is comparable to the age of the system
--- for instance, the values of $n$ and $T$ adopted in the present
calculations imply a flow velocity of $5-50$ km s$^{-1}$ at 10 kpc
for $\dot M=10-100$ $M_{\odot}$ yr$^{-1}$)
and the contribution of nonlinear blobs to the X-ray emission
makes the reduction of $\dot M$
less drastic than that given by the $\dot M \propto r$ model
(Meiksin 1990; Fria\c ca 1993).
These effects, however, do not prevent matter from being removed from the
flow, and the resulting $\phi$ is $\approx 0.2-0.3$.
Therefore, mass removal from the \cf\ 
limits the energetic input of AH, and the luminosities given
in the previous section should be reduced by a factor $\phi$.
Even if AH cannot be invoked as the sole mechanism powering the optical
line emission of class II filaments,
it could be, however, an important ingredient
to explain the emission coming from low ionization lines in these systems.
The models with nonlinear heating and high
AH efficiencies exhibit a strong [OI]$\lambda$6300 emission, and,
therefore, AH
could be an additional component together with other mechanisms for producing
most of the emission in H$\alpha$
and higher ionization lines (as [NII]$\lambda$6583).
The combination of two distinct heating mechanisms, each one being responsible
for certain emission lines and/or explaining one class of filaments,
has allowed to build hybrid models for optical filaments, which have been
particularly succesful at reproducing the observations of optical filaments.
Shocks models tend to reproduce the locus in the line-ratio diagram
of class II filaments, although the energetic requirements are stringent.
A hydrid model could, therefore, combine shocks and an additional 
heating source which explains the higher ionization class I filaments.
In this way, Crawford \& Fabian (1992) invoke shocks between cold clouds
to explain the class II filaments (and produce their [OI] emission)
and mixing layers to account for the class I filaments with
lower H$\alpha$ luminosities and higher levels of ionization 
(higher [NII]$\lambda$6583/H$\alpha$ ratios).
Other additional heating sources are
magnetic reconnection (Jafelice \& Fria\c ca 1996),
and Alfv\'en heating (this work).
Jafelice and Fria\c ca (1996) suggest a hydrid model involving
magnetic reconnection (with an efficient producion of [OI] emission).
and self-absorbed mixing layers with high flux and low temperature for the
incident radiation (producing most of the H$\alpha$ and [NII] emission).
Whether AH is at work together with shocks or with mixing layers, it
efficiently supplies [OI] emission.
For a representative 
$\tilde L_{H\alpha} = 1\times 10^{41}\phi$ erg s$^{-1}$
of model E, one obtains a [OI]$\lambda$6300 luminosity
$\approx 2\times 10^{41}\phi$ erg s$^{-1}$ (see Fig.~2).
Given the average ratio [OI]$\lambda$6300/H$\alpha = 0.25$ of the optical
filaments, AH could account for the [OI]$\lambda$6300 emission of
systems with $\tilde L_{H\alpha}$ up to 
$\approx 8\times 10^{41}\phi$ erg s$^{-1}$.
The AH keeps the gas relatively warm,
giving rise to strong emission of [OI]$\lambda$6300.
We note that the energetic requirements concerning the [OI] emission
of the more luminous systems, like Perseus,
with $\tilde L_{H\alpha}=2.6\times 10^{42}$ erg s$^{-1}$
(considering a H$\alpha$ luminosity of $4.7\times 10^{42}$ erg s$^{-1}$
(Heckman et al. 1989) and a cooling flow rate of 183 M$_{\odot}$ yr$^{-1}$
(Fabian 1994)), are not satisfied by our model even in the most favorable case. 
For these extreme systems, an additional heating mechanism is required.
Whereas there is undoubtedly a connection between cooling flows and
optical filaments, this connection is troubled by some facts, and
our model can shed light in this issue.
In the first place, 
while there is a strong correlation between the H$\alpha$ luminosity
and the mass accretion rate (at a 99.94\% confidence level according to
Heckman et al. 1989), the relation cannot be direct because of the 
large (more than two orders of magnitude) scatter in the relation
$L_{H\alpha}-\dot M$ and the fact that some massive cooling flows 
(e.g., A2029) show no detectable line emission.
The existence of a cooling flow, therefore, is a necessary but not
sufficient condition for detectable line emission.
Secondly, 
except for a few filament systems (e.g., Perseus and A1795)
extending over several tens of kpc, 
the optical line emission is confined to the inner 10 kpc of cooling flows,
whereas the X-ray images of clusters reveal that the cooling gas
is distributed throughout the cooling radius ($\sim 100$ kpc).
The high spatial concentration of optical filaments towards the center
could be explained by a greater magnetic field strength in the central
regions (Soker \& Sarazin 1990) and by a higher level of turbulence
in these regions (Loewenstein \& Fabian 1990; Lowewenstein 1990;
Begelman \& Fabian 1990). As a consequence, an intense Alfv\'en
wave generation is boosted only in the inner cooling flow.
On the other hand, the level of turbulence could be the ``second
parameter" besides mass accretion rate, regulating the luminosity
of optical filament systems. Turbulence, caused, for instance, by
stirring of the central ICM by a galaxy moving through the cluster core,
or by interactions between the ICM and
the relativistic plasma of an extended radio source,
could trigger, via generation and dissipation
of Alfv\'en waves, optical emission lines. 
In this scenario,
the generation of Alfv\'en constitutes a step in the tapping of the
turbulent energy of the ICM into line luminosity.
Additional support for the idea
that turbulence can be crucial for producing the line luminosity
comes from the fact that the kinetic energy flow, determined
from the observed velocity dispersion of the filaments,
correlates much tighter (a scatter of one order of magnitude)
with the H$\alpha$ luminosity (Heckman et al. 1989) than $\dot M$ does.
This indicates that it is the combination of mass flow and 
high levels of turbulence that is relevant for the line luminosity,
rather than a large mass accretion rate alone.
The present calculations are intended to apply to the inner $\sim 10$ kpc
of the \cf s, where the optical filaments are most commonly seen.
The turbulent velocities of the \cf\ increase toward the center
of the cluster, as inferred from the increase of the line widths near
the center (Heckman et al. 1989). 
Since the turbulent energy of the \cf\ is the energy source of the Alfv\'en
waves, 
it is only in the inner $\sim 10$ kpc that the density of turbulent energy
is high enough to make AH efficient.
In this region, blobs condensating out of the \cf\ are kept
warm enough by AH to produce strong line emission.
Blobs cooling down in the outer regions are subject to low levels of AH
and cool quiescently without noticeable optical
emission (similarly to model A).  
In this way, the AH mechanism is consistent with the fact
that line emission does not extend to $\sim 100$ kpc.
A further issue raised by the results of our models concerns the growth of
thermal instabilities in cooling flows.
It has been suggested that a number of processes could inhibit
the growth of thermal instabilities in the presence of gravitational fields
and background flow in cooling flows (Malagoli et al. 1987;
Balbus 1988; Balbus \& Soker 1989; Tribble 1989; Loewenstein 1989;
Brinkmann et al. 1990;
Hattori \& Habe 1990; Yoshida et al. 1991; 
Malagoli et al. 1990; Reale et al. 1991).
The fact that one of our models 
(with nonlinear heating, $\beta_{on}=3$, and $\zeta=1$)
has reached thermal stability shows that
also Alfv\'en heating can constitute a mechanism for suppressing
strong growth of thermal instabilities in cooling flows.
On the other hand,
Gon\c calves et al. (1993a; 1996a) have considered the possibility that
the broad line regions of quasars are formed via thermal instability
in the presence of AH.
They have investigate three damping mechanisms of Alfv\'en waves
-- resonance surface, nonlinear, and turbulent --,
and found that AH could establish a stable two-phase medium
in the broad line regions of quasars.
The thermal stability of the model mentioned above suggests that, 
also in the case of the cooling flow medium,
AH could induce a stable two-phase equilibrium.
\acknow{
One of the authors (D.R.G.) would like to thank the brazilian agency
FAPESP for support, and the other authors
(A.C.S.F, L.C.J., V.J.P. and R.O.) would like to thank the brazilian agency
CNPq for partial support.
We thank the anonymous referee whose comments helped us to improve
the paper significantly.}
\begref {References}
\ref
Allen S.W. et al., 1992, MNRAS 259, 67
\ref
Balbus S.A., 1988, ApJ 328, 395
\ref
Balbus S.A., Soker N., 1989, ApJ 341, 611
\ref
Baum S.A., 1992. In: Fabian A.C. (ed.) Clusters and Superclusters of
Galaxies. Kluwer, Dordrecht, p.171
\ref
Begelman M., Fabian A.C., 1990 MNRAS, 183, 367
\ref
B\"ohringer H., Fabian A.C., 1989, MNRAS 237, 1147
\ref
Brinkmann W., Massaglia S., M\"uller E., 1990, A\&A 237, 536
\ref
Chin Y.C., Wentzel D.G., 1972, Astrop. Space Science 16, 465
\ref
Cowie L.L., Hu E., Jenjins, York D., 1983, ApJ 272, 29
\ref
Crawford C.S., Fabian A.C., 1992, MNRAS 259, 265
\ref
Crawford C.S., Edge A.C., Fabian A.C., Allen S.W., B\"ohringer H.,
 McMahon R.s W., 1995, MNRAS 274, 75
\ref
David L.P., Bregman J.N., 1989, ApJ 337, 97
\ref
David L.P., Bregman J.N., Seab C.G., 1988, ApJ 329, 66
\ref
De Groot J.S., Katz J.I., 1973, Phys. Fluids 16, 40
\ref
Donahue M., Voit G.M., 1991, ApJ 381, 361
\ref 
dos Santos L.C., Jatenco--Pereira V., Opher R., 1993, ApJ 410, 732
\ref
Eilek J.A., Henriksen R.N., 1984, ApJ 277, 820
\ref
Fabian A.C., 1994, ARAA 32, 377
\ref
Fabian A.C., Nulsen P.E.J., Canizares C.R., 1984, Nat 311, 733
\ref
Fabian A.C., Nulsen P.E.J., Canizares C.R., 1991, A\&AR 2, 191
\ref
Fria\c ca A.C.S., 1986, A\&A 164, 1
\ref
Fria\c ca A.C.S., 1993, A\&A 269, 145
\ref
Fria\c ca A.C.S., Terlevich R.J., 1994.
In: Tenorio-Tagle G. (ed.) Violent Star Formation from 30 Douradus to QSOs.
Cambridge University Press, Cambridge, p. 424
\ref
Fria\c ca A.C.S., Terlevich R.J., 1996, MNRAS, submitted
\ref
Ge J.P., Owen F.N. 1993, AJ 105, 778
\ref
Gon\c calves D.R., Jatenco--Pereira V., Opher R., 1993a, ApJ 414, 57
\ref  
Gon\c calves D.R., Jatenco--Pereira V., Opher R., 1993b, A\&A 279, 351
\ref
Gon\c calves D.R., Jatenco--Pereira V., Opher R., 1996a, ApJ 463, 489
\ref
Gon\c calves D.R., Jatenco--Pereira V., Opher R., 1996b, ApJ, submited
\ref
Grevesse N., Anders E., 1989. In: Waddington C.J. (ed.) Cosmic Abundances of
Matter. Am. Inst. Phys., New York, p. 183
\ref
Gruenwald R.B., Viegas S.M., 1992, ApJS 78, 153
\ref
Hattori M.,  Habe A., 1990, MNRAS 242, 399
\ref
Heckman T.M., 1981, ApJ 250, L59
\ref
Heckman T.M., Baum S.A., Van Breugel W.J.M., McCarthy P., 1989, ApJ 338, 48
\ref
Hollweg J.V., 1986, J. Geophys. Res 91, 411
\ref
Hollweg J.V., 1987. In: Proc. 21st ESLAB Symp. 
on Small--Scale Plasma Processes (Esa SP -- 275), p. 161
\ref
Hu E.M., Cowie L.L., Wang Z., 1985, ApJS 59, 447
\ref
Jafelice, L.C., Fria\c ca, A.C.S., 1996, MNRAS, in press
\ref
Jatenco--Pereira V., Opher R., 1989a, A\&A 209, 327
\ref
Jatenco--Pereira V., Opher R., 1989b, MNRAS 236, 1
\ref  
Jatenco--Pereira V., Opher R., 1989c, ApJ 344, 513
\ref
Johnstone R.M., Fabian A.C., Nulsen P.E.J., 1987, MNRAS 224, 75
\ref
Kaburaki O., Uchida Y., 1971, PASJ 23, 405
\ref
Kainer S., Dawson J., Coffey T., 1972, Phys. Fluids 15, 2419
\ref
Lagage P.O., Cesarsky C.J., 1983, A\&A 125, 249
\ref
Loewenstein M., 1989, MNRAS 238, 15
\ref
Loewenstein M., 1990. In: D.J. Hollenbach \& H.A. Thronson Jr. (eds.)
The Interstellar Medium in External Galaxies. 
Nasa Conference Publication 3084, Washington, p. 191
\ref
Loewenstein M., Fabian A.C., 1990, MNRAS 242, 120
\ref
Malagoli A., Rosner R., Bodo G., 1987, ApJ 319, 632
\ref
Malagoli A., Rosner R., Fryxell B., 1990, MNRAS 247, 367
\ref 
Matthaeus W.H., Goldstein M.L., 1982 , J. Geophys. Res. 87, 6011
\ref
Meiksin A., 1990, ApJ 352, 466
\ref
Nulsen P.E.J., Stewart G.C., Fabian A.C., 1984, MNRAS 208, 185
\ref
Opher R., Pereira V.J.S., 1986, Astrophys. Lett. 25, 107 
\ref
Reale F., Rosner R., Malagoli A., Peres G., Serio S., 
1991, MNRAS 251, 379
\ref
Rephaeli Y., Gruber D.E., 1988, ApJ 333, 133
\ref
Soker N., Sarazin C.L., 1990, ApJ 348, 73
\ref
Taylor G.B., Barton E.J., Ge J.P., 1994, AJ 107, 1942
\ref
Thomas P.A., Fabian A.C., Nulsen P.E.J., 1987, MNRAS 228, 973
\ref
Tribble P.C., 1989, MNRAS 238, 1
\ref
Tu C.Y., Pu Z.Y., Wei F.S., 1984, J. Geophys. Res.  89, 9695
\ref 
Uchida Y., Kaburaki O., 1974, Solar Physics 35, 451
\ref
Voit G.M., Donahue M., 1990, ApJ 360, L15
\ref
Voit G.M., Donahue M., Slavin J.D., 1994, ApJS 95, 87
\ref
Wentzel D.G., 1974, Solar Physics 39, 129
\ref
Yoshida T., Hattori M., Habe A., 1991, MNRAS 248, 630
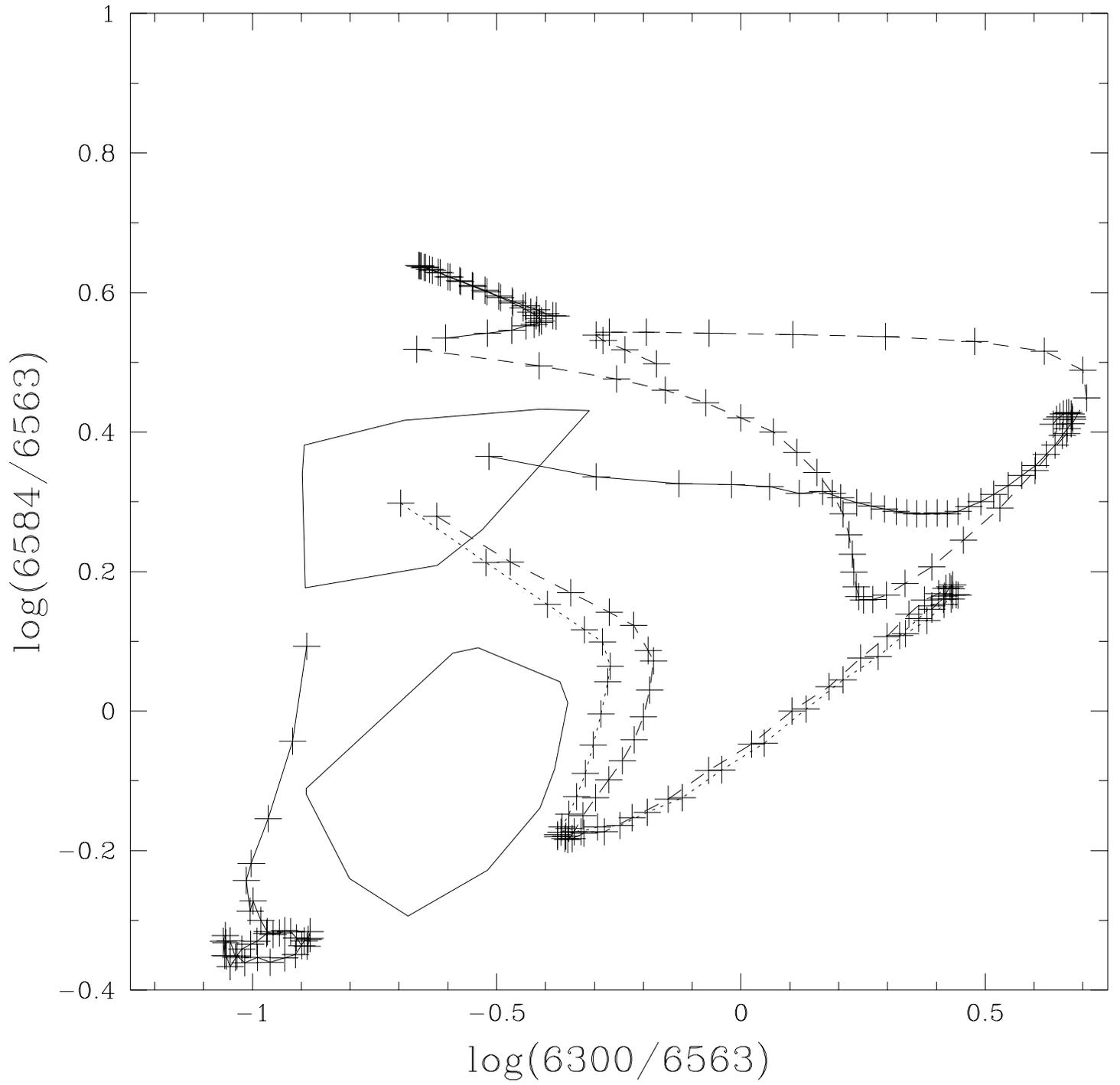
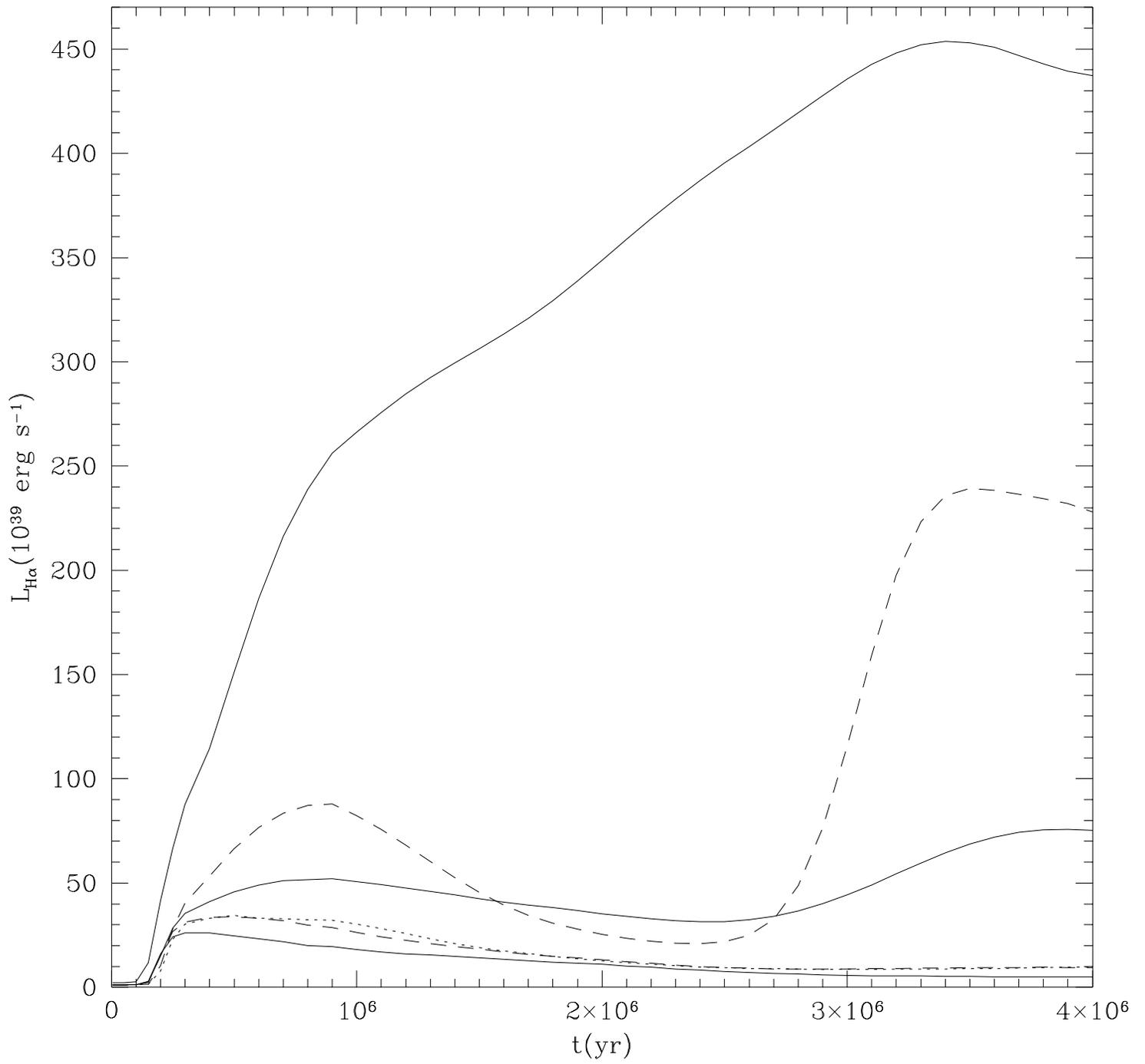
\bye